# The effects of halogen elements on the opening of an icosahedral $B_{12}$ framework

Liang-Fa Gong,[a] Wei Li,[a] Edison Osorio,[b] Xin-Min Wu,[a] Thomas Heine [c,d] and Lei Liu [d*†]

[a] National Demonstration Center of Chemistry and Chemical Engineering Experimental Teaching, School of Chemical Engineering, Beijing Institute of Petrochemical Technology, Beijing 102617, China

[b] Departamento de Ciencias Básicas, Fundación Universitaria Luis Amigó, SISCO, Transversal 51A # 67B 90, Medellín, Colombia

[c] Wilhelm-Ostwald-Institut für Physikalische und Theoretische Chemie, Universität Leipzig, Linnéstr. 2, 04103 Leipzig, Germany

[d] Department of Physics & Earth Sciences, Jacobs University Bremen, Campus Ring 1, 28759 Bremen, Germany

[†] Current address: Max Planck Institute for Polymer Research, Ackermannweg 10, 55128 Mainz, Germany

E-mail: Lei Liu, liulei3039@gmail.com

**Abstract:** The fully halogenated or hydrogenated $B_{12}X_{12}^{2-}$ (X = H, F, Cl, Br and I) clusters are confirmed to be icosahedral. On the other hand, the bare $B_{12}$ cluster is shown to have a planar structure. A previous study showed that a transformation from an icosahedron to a plane happens when 5 to 7 iodine atoms are substituted (*Chem. Eur. J.* 18 (41), 13208-13212). Later, the transition was confirmed to be seven iodine substitutions based on an infrared spectroscopy study (*Chem. Phys. Lett.* 625, 48–52). In this study, we investigated the effects of different halogen atoms on the opening of the $B_{12}$ icosahedral cage by means of density functional theory calculations. We found that the halogen elements do not have significant effects on the geometries of the clusters. The computed IR spectra show similar representative peaks for all halogen substituted clusters. Interestingly, we found a blue-shift in the IR spectra with the increase in the mass of the halogen atoms. Further, we compared the Gibbs free energies at different temperatures for different halogen atoms. The results show that the Gibbs free energy differences between *open* and *close* structures of $B_{12}X_7^-$ become larger when heavier halogen atoms are present. This interesting finding was subsequently investigated by energy decomposition analysis.

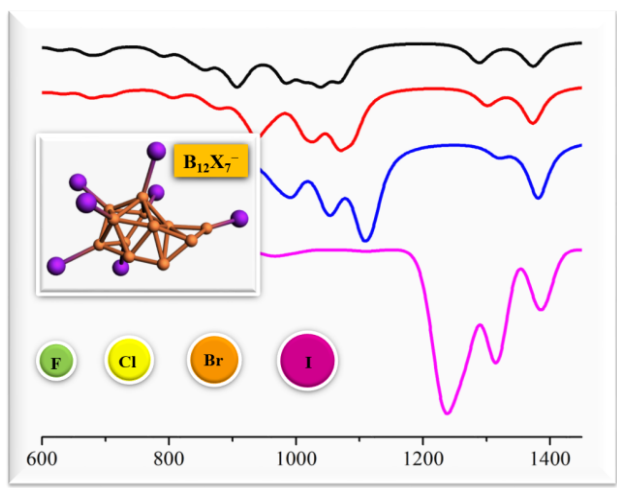

The effects of halogen elements on the opening of an icosahedral $B_{12}$ framework have been investigated by density functional theory calculations and infrared spectroscopy.

**Introduction**

Together with other elements, boron exists in different forms of compounds. Examples of these compounds are boranes (or borohydrides),[1] carboranes,[2] and metallacarboranes.[3] Their peculiar electronic structures and unusual chemical bonds result in a variety of interesting applications in the field of medicine[4] and in the production of coordination polymers,[5] liquid crystals,[6] ionic liquids[7] as wells luminescent materials.[8] Typically, boron compounds show three-dimensional (3D) deltahedra structures with multiple B–B bonds (i.e. multiple-center two-electron bonds).[9–11] Among them, a heavily studied example is the fully iodine or hydrogen substituted $B_{12}I_{12}^{2-}$ or $B_{12}H_{12}^{2-}$ clusters.[12] These clusters have been confirmed to icosahedral (a central icosahedral $B_{12}$ cage with substitutions in the periphery). Like carbon atoms, boron itself can also form atomic clusters, say boron clusters. Unlike boron compounds in deltahedra structures, small pure boron clusters without further substituents are essentially planar.[13–15] Until now, the biggest pure boron cluster which is confirmed to be planar is the $B_{40}^-$ cluster, of which the global minimum structure is a planar structure with two adjacent hexagonal holes.[16] Note that many planar boron clusters have hexagonal holes in their most stable structures, and such geometric characteristic was attributed for their high stability.[17]

An interesting question is at what degree of substitution the icosahedral $B_{12}$ cage opens when a fully saturated $B_{12}I_{12}^{2-}$ cluster undergoes a step by step stripping of the iodine substituents. Previously, a complete series of partly iodinated $B_{12}I_n^{x-}$ (with $n$ = 1 to 12, $x$ = 1 or 2) was successfully generated in the gas-phase and was studied by mass spectroscopy.[18] Together with theoretical calculations, it was shown that the most stable structures for $n \geq 8$ remain the icosahedral $B_{12}$ cages. The planar structures are preferred for $n \leq 4$. The transition between the icosahedral and planar structures is found be $n$ = 5 to 7 depending on the temperatures.[9] Later, a combined experimental and theoretical infrared (IR) spectroscopy study was carried out to narrow down the region of transition.[19] The results showed that the simulated IR spectra matches the experimental IR spectra based on the icosahedral structures for $n \geq 8$. The simulated IR spectra of *open*-$B_{12}I_7^-$ cluster showed a certain degree of correspondences to the experimental IR spectra. These results indicated that the icosahedral $B_{12}$ cage opens at $B_{12}I_7^-$ and most probably the experimentally obtained $B_{12}I_7^-$ clusters are the mixture of low-lying isomers of both *close* and *open* structures.

Until now, all studies are based on the iodine substitutes. To the best of our knowledge, there is no previous study investigating the effects of the different halogen atoms on the opening of the icosahedral $B_{12}$ cage. To this end, we performed density functional theory (DFT) calculations in this study. We firstly investigated the effects on the geometries and the IR spectra by changing the substitutions from fluorine (F), chlorine (Cl) and bromine (Br) to iodine (I). Subsequently, we computed the Gibbs free energy difference ($\Delta G$) between the *close-* and *open-*$B_{12}X_7^-$ clusters (X= F, Cl, Br and I), and the electronic energy ($\Delta E$) contributions are studied by the energy decompositions analysis (EDA).

**Computational details**

All DFT calculations were carried out with the Gaussian 09 package.[20] All structures were fully optimized at the PBE0 level of theory,[21] with the basis set of def2-TZVP.[22] Within the calculations, the effective core basis set of LanL2DZ was assigned for iodine atom.[23] The harmonic frequency calculations were performed at the same level of theory to characterize the nature of stationary points, i.e. no imaginary frequencies were found for all optimized structures. The initial structures were taken from previous studies by replacing I with F, Cl and Br.[9,19] The PBE0 functional was benchmarked as a reliable DFT method for boron systems in terms of both geometries and energetics.[23] The combination of a PBE0 functional with a def2-TZVP basis set has been widely used to study the electronic structures of boron clusters.[11,17,24] The bonding interactions corresponding to the formation of *close-* and *open-*$B_{12}X_7^-$ have been analyzed with the EDA scheme.[25] These calculations were performed at PBE0 level with the basis set of TZ2P.[26] The scalar relativistic effects for the iodine atom were incorporated by zeroth order regular approximation (ZORA).[27] All EDA calculations were performed by employing ADF software.[28]

**Results and discussions**

All optimized structures, $B_{12}X_{12}^{2-}$, *close-*$B_{12}X_7^-$, *open-*$B_{12}X_7^-$ and bare $B_{12}$ clusters, are depicted in **Figure 1** and the selected distances are provided in **Table S1-S3**. The most stable structures of $B_{12}X_{12}^{2-}$ show icosahedral configurations and have the symmetry of $I_h$. In these structures, all B-B bond lengths are the same (~ 1.80 Å), indicating almost no effects of different halogen atoms on the geometry of the icosahedral $B_{12}$ cage. The B-X bond lengths gradually increase from 1.38 Å ($B_{12}F_{12}^{2-}$) to 2.18 Å ($B_{12}I_{12}^{2-}$) which is in the same order to the van der Waals atomic radius of halogen atoms. The bare $B_{12}$ cluster without any halogen substitutions has a $C_{3v}$ symmetry and the B-B bond length can be categorized into three groups: 1) shortest B-B bond length of about 1.55 Å for B6 and B7; 2) medium B-B bond length about 1.60 for B7-B4, B7-B5 and B12-B5; and 3) longest B-B bond length for B4-B5, which is about 1.80 Å. The *close* structures of $B_{12}X_7^-$ overall remain the icosahedral $B_{12}$ framework but with some changes in the B-B bond length distributions. For example, the bond lengths of B1-B2 and B1-B4(B5) are 1.60 and 1.67 Å, respectively. The bond lengths between B2 and B4 (B6), B3 and B4, B5 and B6 are 1.75 Å while the bond length for B4-B5 is only 1.58 Å. In general, the B-B bond lengths in the lower part of the $B_{12}X_7^-$ clusters (Figure 1b) are longer than the upper part. For example, the bond length of B10-B12 is 1.74 Å, and bond lengths of B7(B8)-B12 and B9(B11)-B12 are more than 1.80 Å. This might be the consequence of the Pauli repulsion between the halogen atoms and there are more halogen atoms in the lower part than the upper part (five halogen atoms in the lower part while only two in the upper part). Note that we do not find significant effects of the different halogen atoms on the B-B bond length distributions, and all *close*-$B_{12}X_7^-$ clusters show similar B-B bond lengths (for more details, see Table S1). The most stable *open* structures of $B_{12}X_7^-$ are very complicated as shown in Figure 1c. In general, they show both features of the icosahedral $B_{12}X_{12}^{2-}$ and the bare $B_{12}$ clusters. For example, structures with a vertex boron atom (i.e. B2) surrounding by five boron atoms, similar to that in the icosahedral $B_{12}X_{12}^{2-}$ cluster. On the other hand, there are also boron atoms locating on the boarder similar to a bare $B_{12}$ cluster, such as B11 and B12. The bond lengths between the vertex B2 atom and its neighbors are about 1.80 Å, which is again similar to the vertex boron atoms in the icosahedral $B_{12}X_{12}^{2-}$ clusters. The bond lengths between the boron atoms on the boarder vary from 1.6 to 2.0 Å. For example, the bond length of B11-B12, in which there are no halogen substitutions, is only 1.58 Å. However, the bond length of B5-B9 is more than 2.0 Å.

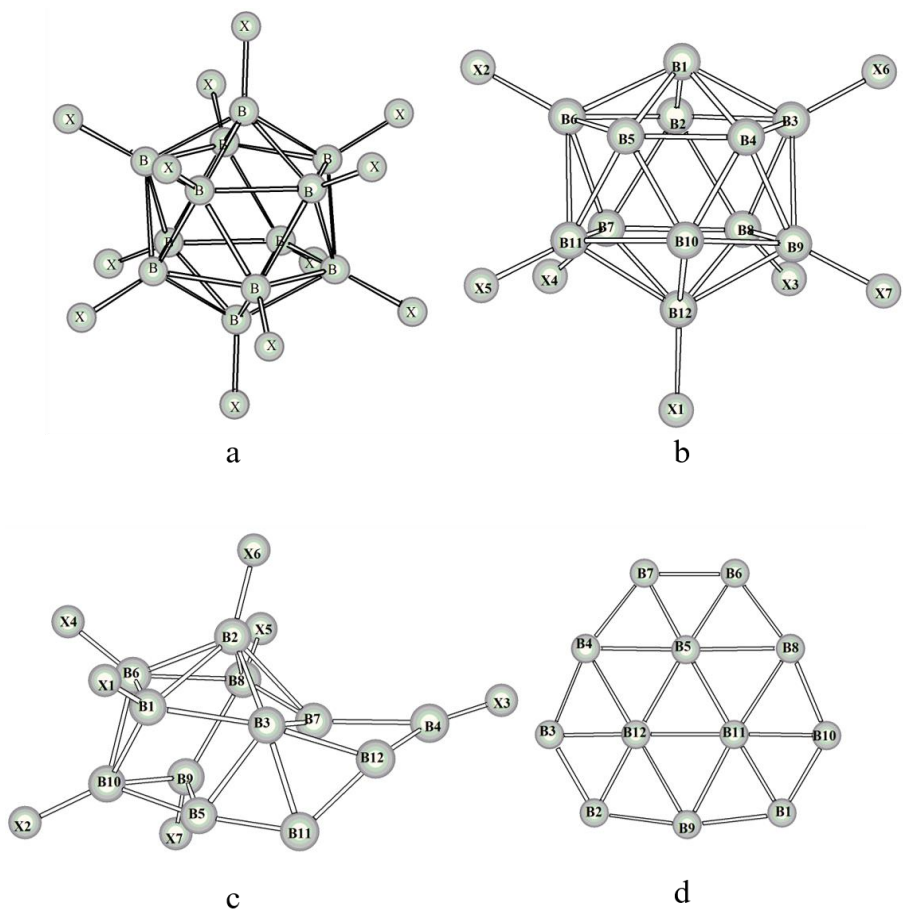

**Figure 1**. The most stable structures of $B_{12}X_{12}^{2-}$ (**a**: $I_h$), *close*-$B_{12}X_7^-$ (**b**: $C_s$), *open*-$B_{12}X_7^-$ (**c**: $C_1$) and $B_{12}$ (**d**: $C_{3v}$) clusters (X= F, Cl, Br and I).

It was previously shown that the *open* and *close* structures have different features in the IR spectra.[19] For example, the *close* (or icosahedral) structures often show adsorption peaks below 1000 cm$^{-1}$ and the overall IR profile is quite simple. The *open* structures often have higher energy absorptions between 1000 and 1300 cm$^{-1}$ and the overall IR profiles are rather complicated. To find out the effects of different halogen atoms on the IR spectra, we performed the IR analysis on both the *close* and *open* structures of $B_{12}X_7^-$, and results are shown in **Figure 2**. In the cases of *close*-$B_{12}X_7^-$, we found two representative peaks, A' and A''. The peak A' corresponds to the deformation motion of the icosahedral $B_{12}$ cage involving the asymmetrical stretching mode of the B-X bonds. The second one, peak A'', corresponds to the breaking motion of the icosahedral $B_{12}$ cage involving the symmetrical stretching mode of the B-X bonds. Note that the IR spectra of the *close* structures of $B_{12}X_7^-$ are generally similar to that of the icosahedral $B_{12}I_{12}^{2-}$ cluster, indicating that the icosahedral $B_{12}$ cage remains intact after the removal of five halogen atoms. This has been also shown by the geometric parameters of the icosahedral $B_{12}I_{12}^{2-}$ and *close*-$B_{12}X_7^-$ clusters, in which the B-B bond lengths are rather close to each other. Interestingly, we found that the positions of the same peaks (A' and A") shift gradually to lower wavenumbers (blue-shift) when the halogen atoms are changed from F to I. For example, the peak A' of the *close*-$B_{12}F_7^-$ located at about 800 cm$^{-1}$ decreases to 680 cm$^{-1}$ when we replaced F by I, while the peak A" shifts correspondingly from 1250 to 1000 cm$^{-1}$. The spectra of the *open*-$B_{12}X_7^-$ generally are complicated and different compared to that of the icosahedral $B_{12}I_{12}^{2-}$ and *close*-$B_{12}X_7^-$ clusters. This is most likely attributed to complex structures of the open forms: low symmetry and different boron coordination. To simplified the discussion, we classified the peaks into two groups, B' and B" (Fig. 2). The group B' consists mainly of the B-B stretching mode of the icosahedral-like B2 and its neighbors (i.e. B1, B3, B6, B7 and B8 in Figure 1c). The second group, peaks B", refers essentially to the B-B stretching mode of the boron atoms on the boarder, such as B11, B12 and B5 in Figure 3c. Similar to that of the *close*-$B_{12}X_7^-$, the positions of the peaks in the case of *open*-$B_{12}X_7^-$ clusters also shift to the lower wavenumbers when we changed the halogen atom from F to I. The largest shift found is for the peaks B' of $B_{12}I_7^-$ which is about 200 cm$^{-1}$. However, we have not found obvious shift of peaks B". This is because the motions of these peaks do not involve halogen atoms while peaks A', A" and B' do have contributions from B-X stretching. When we change the substitutions, the average mass of the groups (i.e. B-X bonds) changes as well. As such, we found the shift in the wavenumbers for such bonds in the IR spectra.

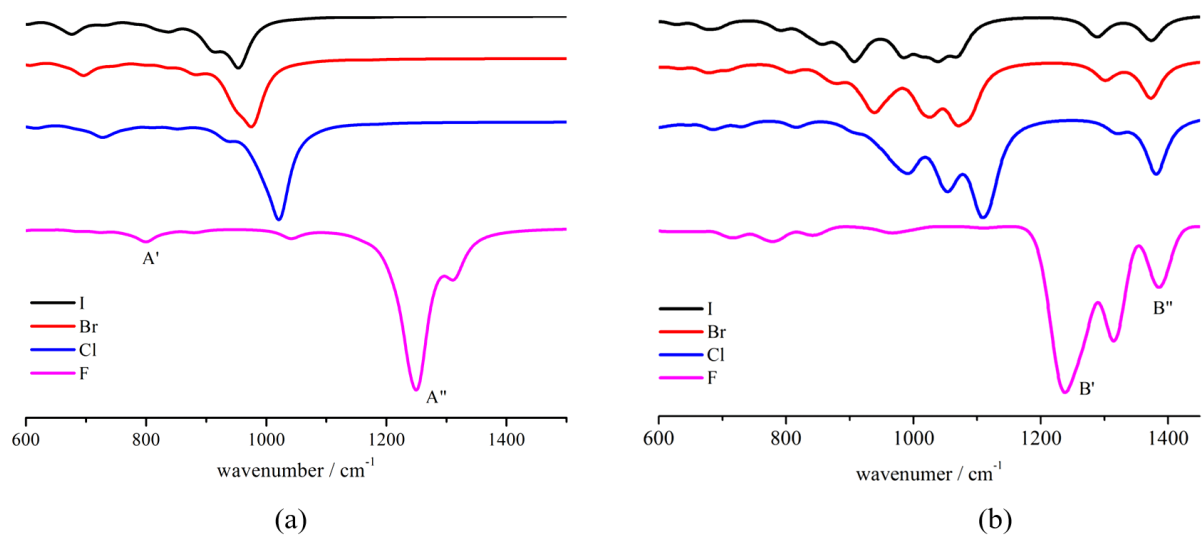

**Figure 2**. simulated IR spetrum of *close*- (a) and *open*- (b) $B_{12}X_7^-$ at PBE0/def2-TZVP (LanL2DZ for I) level of theory. X= F, Cl, Br and I.

The reactivity of the halogenated boron clusters (i.e. $B_{12}X_{12}^{2-}$) is often studied by the electrospray ionization mass spectrometry (ESI-MS).[18] The ions produced by ESI-MS often can capture internal energy in the range of 100 – 200 kJ mol$^{-1}$, which corresponds to temperature of several thousands Kelvin. Therefore, we computed the Gibbs free energies, $\Delta G$, for both *open*- and *close*-B$_{12}$X$_7^-$ at 300, 1500, and 5000 K, through the harmonic approximations for the minimum structures on the potential energy surface at 0 K. The Gibbs free energies were computed by $\Delta G = G_{ico} - G_{ope} = \Delta U + RT - T\Delta S$ ($U$ is the internal energy, $T$ is the temperature, $R$ is the gas constant and $S$ is the entropy), and the data are given in **Table 1**. Previously, it was found that for the B$_{12}$ cluster with seven iodine substituents, the temperature dependence of $\Delta G$ is important,[9] and here we have found the same phenomenon. For example, at 0 and 300 K, the *close*-B$_{12}$I$_7^-$ is favored by 22.3 and 7.8 kcal mol$^{-1}$, respectively, indicating that the B$_{12}$I$_7^-$ cluster has an icosahedral configuration at low temperatures, i.e. in the solid state. However, at 1500 and 5000 K, the *open*-B$_{12}$I$_7^-$ is favored by 8.2 and 55.6 kcal mol$^{-1}$, respectively, indicating that the icosahedral B$_{12}$ cage opens at B$_{12}$I$_7^-$ in the gas-phase. This general trend applies to all halogenated B$_{12}$X$_7^-$ clusters. At low temperature, the *close* form of B$_{12}$X$_7^-$ is favored while *open* forms become dominant when the temperature increases to 1500 and 5000 K. As pointed out by the previous study, at higher temperatures, the *open* structures are favored by the entropy contributions.[9] The important finding in this study is the $\Delta G$ values for different halogen atoms. Let us take the values at 1500 K as an example since this temperature is more realistic considering the thermal energy of the ions in the ESI-MS. For B$_{12}$F$_7^-$, $\Delta G$ is 1.9 kcal mol$^{-1}$. This positive value indicates that the *close* structure is slightly favored over the *open* structure when B$_{12}$ is substituted by seven fluorine atoms. Taking the error of the DFT method into account, we see that B$_{12}$F$_7^-$ might distribute equally as *open* and *close* structures in the gap-phase at 1500 K. In other words, to completely break the icosahedral B$_{12}$ cage of B$_{12}$F$_7^-$ or to have higher concentration of the *open* structures, we need a temperature higher than 1500 K or let the clusters stay for a longer time in the iron trap of the ESI-MS. For B$_{12}$Cl$_7^-$ and B$_{12}$Br$_7^-$, $\Delta G$ is about -5.0 kcal mol$^{-1}$. These larger negative values indicate that a higher concentration of the *open* structures would be obtained in the gas-phase in the cases of Cl and Br. The $\Delta G$ became even more negative in the case of B$_{12}$I$_7^-$, which is -8.2 kcal mol$^{-1}$. This reveals that the *open* forms of the B$_{12}$I$_7^-$ cluster are dominant over the *close* forms at 1500 K. In short, the heavier the halogen atoms presented in the B$_{12}$X$_7^-$ cluster, the lower the temperature needed to open the icosahedral B$_{12}$ cage. This is most probably because of the pulling manner of halogen atoms in the breaking B-B bonds. Heavier elements have stronger strength to pull the boron atoms and is easier to break the B-B bonds.

**Table 1**. Calculated Gibbs free energy difference between *open-* and *close-*$B_{12}X_7^-$ at PBE0/def2-TZVP (LanL2DZ for I) level of theory. All values are given in kcal mol$^{-1}$.

|         | $B_{12}F_7^-$ | $B_{12}Cl_7^-$ | $B_{12}Br_7^-$ | $B_{12}I_7^-$ |
|---------|---------------|----------------|----------------|---------------|
| 0 K     | 22.3          | 16.2           | 14.8           | 12.2          |
| 300 K   | 17.8          | 11.8           | 10.4           | 7.8           |
| 1500 K  | 1.9           | -4.2           | -5.8           | -8.2          |
| 5000 K  | -45.5         | -51.9          | -53.8          | -55.6         |

In general, the *open* forms of $B_{12}X_7^-$ cluster became more stable over the *close* forms from F to I (Table 1). At 0 and 300 K, all $\Delta G$ values were positive, indicating that the *close* forms of $B_{12}X_7^-$ are favored for all halogen atoms. However, the $\Delta G$ vales show similar trend to that computed at higher temperatures, *i.e.* 1500 and 5000 K. Interestingly, the difference in $\Delta G$ values are almost constant and are independent from the temperatures. Taking $B_{12}Br_7^-$ and $B_{12}I_7^-$ as examples, we observe that the differences between the $\Delta G$ are about 3 kcal mol$^{-1}$ from 0 to 5000 K (the difference between the last two columns in Table 1). This finding indicates that the decrease in the $\Delta G$ values is most likely due to the electronic energy component, instead of the thermal or entropy contributions. Thus, we performed EDA calculations on the ground of electronic energies, $\Delta E$ ($\Delta G$ at 0 K) to study the effects of different halogen atoms on the relative stability between the *open* and *close* forms of $B_{12}X_7^-$. The total electronic energies and their individual contributions of $B_{12}X_7^-$ are summarized in **Table 2**. The calculated intrinsic interaction energies ($E_{total}$) have large negative values, which indicate strong attractions of B-B and B-X pairs. It can be seen that the largest value is given for X = F (ca. -3200 kcal mol$^{-1}$) and the smallest values is for X = I (ca. -2500 kcal mol$^{-1}$). This demonstrates that the iodine complex has the weakest B-B and B-I interactions, while the bonds in the $B_{12}Cl_7^-$ and $B_{12}Br_7^-$ are slightly weaker than that in the fluorine complex. This trend can be also understood by the term of orbital interactions, $E_{orb}$. For X= F, the total energy has the largest value of $E_{orb}$ while X = I has the smallest values, and difference was about 1000 kcal mol$^{-1}$. Thus, we observe a stronger covalent characteristic in $B_{12}F_7^-$ than in $B_{12}I_7^-$. Talking about the absolute values, both $E_{Pauli}$ and $E_{elstat}$ followed the same trend: F> Cl>Br > I. To have a clearer profile, we focus on the energy difference, $\Delta E(open - close)$ for each contribution of the total interactions. In general, the *open* structures are stabilized by Pauli repulsions (negative values) and are destabilized by $\Delta E_{elstat}$ and $\Delta E_{orb}$ (positive values). Since the absolute values of $\Delta E_{Pauli}$ is much bigger than that of $E_{elstat}$, the summation of these two terms gives negative values which are comparable to the absolute vales of $\Delta E_{orb.}$ In short, we conclude that the *open* structures are stabilized by static interactions ($\Delta E_{stat} = E_{Pauli} + E_{elstat}$) and are destabilized by $\Delta E_{orb}$. Changing from F to I, we find that both the $\Delta E_{stat}$ stability and the $\Delta E_{orb}$ destability decrease. However, they decrease at different rates. For example, $\Delta E_{stat}$ is -460.1 kcal mol$^{-1}$ in the case of $B_{12}F_7^-$ and it increases to -281.2 kcal mol$^{-1}$ for $B_{12}I_7^-$. Thus, the stability decreases 177.9 kcal mol$^{-1}$. On the other hand, $\Delta E_{orb}$ is 480.6 kcal mol$^{-1}$ in the case of $B_{12}F_7^-$ and it decreases to 292.2 kcal mol$^{-1}$ for $B_{12}I_7^-$. The corresponding destability decreases 188.4 kcal mol$^{-1}$. Therefore, the difference between the stability from $\Delta E_{stat}$ and the destability from $\Delta E_{orb}$ is 10.5 kcal mol$^{-1}$ when changing from F to I. As such, $\Delta E$ (*open – close*) is 20.4 kcal mol$^{-1}$ for $B_{12}F_7^-$ and it is 11.0 kcal mol$^{-1}$ for $B_{12}I_7^-$.

**Table 2**. The energy decomposition analysis (EDA) at PBE0/ZORA/TZ2P level of theory for *open*- and *close*-$B_{12}X_7^-$ (X= F, Cl, Br and I). All the values are given in kcal mol$^{-1}$.

| cluster | $E_{Pauli}$ | $E_{elstat}$ | $E_{Pauli} + E_{elstat}$ | $E_{orb}$ | $E_{total}$ |
|---|---|---|---|---|---|
| $B_{12}F_7^-$ | | | | | |
| *open* | 13654.51 | -3016.06 | 10638.46 | -13828.08 | -3189.62 |
| *close* | 14378.55 | -3279.99 | 11098.57 | -14308.63 | -3210.05 |
| $\Delta E$ | -724.04 | 263.93 | -460.11 | 480.55 | 20.43 |
| $B_{12}Cl_7^-$ | | | | | |
| *open* | 13129.50 | -2972.65 | 10156.85 | -12892.40 | -2735.55 |
| *close* | 13721.34 | -3230.27 | 10491.07 | -13240.94 | -2749.86 |
| $\Delta E$ | -591.84 | 257.62 | -334.22 | 348.54 | 14.31 |
| $B_{12}Br_7^-$ | | | | | |
| *open* | 12800.46 | -2951.15 | 9849.32 | -12456.66 | -2607.35 |
| *close* | 13363.83 | -3204.90 | 10158.93 | -12779.53 | -2620.60 |
| $\Delta E$ | -563.37 | 253.75 | -309.61 | 322.87 | 13.25 |
| $B_{12}I_7^-$ | | | | | |
| *open* | 12417.68 | -2879.42 | 9538.25 | -12006.46 | -2468.21 |
| *close* | 12948.99 | -3129.50 | 9819.49 | -12298.65 | -2479.16 |
| $\Delta E$ | -531.31 | 250.08 | -281.24 | 292.19 | 10.95 |

**Conclusions**

In this study, we investigated the effects of different halogen atoms (F to I) on the opening of the icosahedral $B_{12}$ cage by means of DFT calculations. We firstly investigated the effects on the geometric parameters as well as the IR spectra by changing the substitutions from F to I. Subsequently, we computed the Gibbs free energy differences ($\Delta G$) between the *open-* and *close-*$B_{12}I_7^-$ clusters and the electronic energy contributions were studied by the energy decompositions analysis (EDA). We found that the halogen atoms do not have significant effects on the geometry of the structures. Moreover, the computed IR spectra showed similar representative peaks for all halogen atoms. Interestingly, while increasing the mass of the halogen atoms, we found a blue-shift in the IR spectra. The largest shift found was for the peaks B' (the vibrational mode of the icosahedral-like boron atoms) of *open*-$B_{12}X_7^-$ which was about 200 cm$^{-1}$ in the case of iodine substitution. Lastly, we compared the Gibbs free energies at different temperatures for different halogen atoms. It was shown that the *open* structures of $B_{12}X_7^-$ became more stable over the *close* structures when heavier halogen atoms were present. EDA results show that this general trend is because the stability from static interactions ($\Delta E_{stat}$) and destability from orbital interactions ($\Delta E_{orb}$) decrease in different rates.

**Notes**

The authors declare no competing financial interest.

**Acknowledgement**

We thank Chemical Design Laboratory of National Demonstration Center of Chemistry and Chemical Engineering Experimental Teaching, Beijing Institute of Petrochemical Technology for allocation of computational resources.